\documentclass[aps,pra,twocolumn,groupedaddress,showpacs]{revtex4}

\usepackage{amsmath,amssymb}
\usepackage{graphicx}

\def\rnum#1{\expandafter{%
\romannumeral #1}}
\def\Rnum#1{\uppercase\expandafter{%
\romannumeral #1}}

\newcommand{\bol}[1]{\boldsymbol #1}

\begin{document}


\title{Spontaneous rotation in one-dimensional systems of cold atoms}


\author{Akiyuki Tokuno$^{1,2}$ and Masahiro Sato$^{3}$}
\affiliation{
$^1$Department of Physics, Tokyo Institute of Technology,
Oh-okayama, Meguro-ku, Tokyo 152-8551, Japan \\
$^2$Institute for Solid State Physics, University of Tokyo, Kashiwa
227-8581, Japan \\
$^3$Condensed Matter Theory Laboratory, RIKEN, Wako, Saitama 351-0198, Japan}


\date{\today}

\begin{abstract}
 We theoretically study 
 harmonically trapped one-dimensional Bose gases (e.g., ${}^7{\rm Li}$,
 ${}^{23}{\rm Na}$, ${}^{39}{\rm K}$, ${}^{87}{\rm Rb}$, etc.) with multibands
 occupied, focusing on effects of higher-energy bands. 
 Combining the Ginzburg-Landau theory with the bosonization
 techniques,
 we predict that the repulsive interaction between higher-band bosons 
 and the quantum fluctuation can induce the ground state with a finite 
 angular momentum around the trapped axis. 
 In this state, the $Z_2$ reflection symmetry
 (clockwise or anticlockwise rotations) is spontaneously broken. 
\end{abstract}


\maketitle


\section{Introduction}
For the last decade, cold-atom systems~\cite{Pethick,Pitaevskii,CW,Leggett} 
have been vigorously studied 
and experimental techniques of controlling them have been greatly
developed. As a result, we can now tune various parameters of 
cold-atom systems, even including dimensionality. 
These developments have enabled us to encounter 
several phenomena: For instance,
superfluid-insulator transitions on optical lattices, 
dynamics of Bose-Einstein condensations (BECs), BCS-BEC crossover 
in Fermi gases, etc. They are all difficult to be realized 
within traditional experiments using electronic systems in solids.

Among various new fields of cold-atom physics, 
in this paper, we focus on one-dimensional (1D) trapped Bose gases
~\cite{FZ,TOZD,GVL,Olshanii,Cazalilla}: 
For example, polarized atoms ${}^7{\rm Li}$, ${}^{23}{\rm Na}$, 
${}^{39}{\rm K}$, ${}^{87}{\rm Rb}$, etc. 
One-dimensional quantum systems have been shown to
exhibit unusual interesting features due mainly to effects 
of strong quantum fluctuations. For example,
it is widely known that BECs never occur even at zero temperature in 
1D repulsively interacting Bose gases in the thermodynamic limit. 
Instead, a Tomonaga-Luttinger liquid (TLL) state is
predicted to appear. Actually TLL and Tonks-Girardeau (TG) 
gas~\cite{Girardeau} $-$ the hard-core interacting limit
of TLL $-$ behaviors have been observed in recent experiments 
of 1D cold Bose gases.~\cite{KWW,PWM}
 


\begin{figure}[bp]
 \begin{center}
  \includegraphics[scale=0.6]{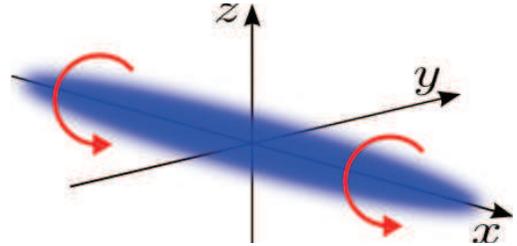}
  \caption{One-dimensional trapped Bose gas confined 
  along the $x$ axis. In the text, we predict that the Bose gas 
  spontaneously circulates along the arrow (or along the opposite
  direction of the arrow) under a certain
  condition.}
  \label{trapped_gas}
 \end{center}
\end{figure}

One-dimensional confined systems in cold-atom experiments often take 
a cigar shape as shown in Fig.~\ref{trapped_gas}. 
This shape is of course caused by a 2D ($y$-$z$ plane) potential, 
which strongly restricts the in-plane dynamics and 
makes the radial energy spectrum discrete, i.e., yields a multi-band structure. 
The dimensionality of the cigar-shaped systems 
is known to be determined by the 3D s-wave scattering length $a$ 
between atoms and the particle density $\rho$ per 1D unit length. 
For $a\rho\ll1$ the system may be regarded as a purely 1D gas, namely,
almost all of the atoms are in the lowest band. On the other hand, 
for the opposite case $a\rho\gg1$, 3D nature strongly survives. 
In the former case, the low-energy physics is characterized 
by a single dimensionless parameter $\gamma^{-1}=a_{1D}\rho$, 
where $a_{1D}$ is the effective 1D
scattering length.~\cite{Olshanii} A small
$\gamma$ corresponds to the weakly interacting regime in which
low-energy properties can be captured by mean-field
(Gross-Pitaevskii-type) theory, whereas for a large $\gamma$, 
effects of the interaction is considerably strong and 
thus the TLL or TG-gas behaviors clearly emerge.  

Thanks to the recent studies of purely 1D atomic gases, 
their understanding has been greatly deepened. 
However, 
there still exist important issues in 1D confined systems: 
One of them is what happens if atoms
occupy not only lowest band but also higher ones. 
As yet, effects of higher bands have not been well discussed well.  
Since multi bands may be regarded as an internal degrees of freedom 
such as electron spins, the system is expected to contain rich
physical properties. In this paper, we will provide an answer to this
problem. To this end, we consider a simple 3D single-component 
Bose gas system with a repulsive interaction and a 2D harmonic
potential. As we will explain, due to the potential, 
each energy band takes an angular momentum $l$: the lowest band has 
$l=0$, the second lowest ones have $l=\pm 1$ and are hence doubly
degenerate, and so on. A positive (negative) $l$ 
corresponds to atoms rotating clockwise (anticlockwise) around 
the confinement axis. 
For simplicity, we will concentrate on the case where 
only the lowest band and the degenerate second lowest ones 
(totally three bands) are occupied. 
In this situation, it is shown that at least if the 2D harmonic potential
is sufficiently stronger than the interaction between atoms, 
the particle densities of the $l=\pm1$ bands become imbalanced and 
the ground state spontaneously rotates with a finite
angular momentum as shown in Fig.~\ref{trapped_gas}.
We will explain in detail the mechanism of this 
spontaneous rotation phenomenon.


This paper is organized as follows. 
First, we define the 3D Hamiltonian of the Bose atoms
confined by a 2D harmonic potential in Sec.~\ref{model}. 
Then, we reduce the 3D Hamiltonian to the 1D effective one 
in order to describe the low-energy physics.
In Secs.~\ref{analysis1} and \ref{analysis2}, 
based on the reduced Hamiltonian, we analyze the low-energy physics 
of the trapped bosons. Employing the Ginzburg-Landau (GL) theory 
and the bosonization techniques, we show that the spontaneous rotation 
as in Fig.~\ref{trapped_gas} can occur under certain, realistic
conditions. The final Section~\ref{summary} is devoted to 
the summary and the discussions for our results. 
In Appendix~\ref{sym_argument}, we summarize the symmetries of 
our Bose gas system, which play important roles in the analyses of 
Secs.~\ref{analysis1} and \ref{analysis2}.

\section{Bose Gas and Effective Hamiltonian}\label{model}
First, we introduce a 3D Bose gas in the presence of 
a 2D harmonic potential, whose Hamiltonian  is defined as 
\begin{equation}
 H
 =\int\!\!d\vec{r}
  \left[
   \psi^{\dagger}
   \left(
    -\frac{\hbar^2\nabla^2}{2m}
    -\mu+V({\vec r})
   \right)
   \psi
   +\frac{U}{2}{\cal D}^2+\cdots
  \right], \label{3DHamiltonian}
\end{equation}
where $\psi({\vec r})$ and ${\cal D}=\psi^\dag\psi$ are 
the annihilation field and the density operator 
of the bosons, respectively. Positive parameters $m$, $\mu$ and $U$ 
respectively mean the mass of bosons, the chemical potential and
the repulsive coupling constant. The 3D s-wave scattering length $a$ is
related to $U$ via $U=4\pi\hbar^2a/m$. Three or more-body interactions
are assumed to be quite small compared with $U$. The 2D confinement
potential $V$ is a harmonic type:
$V({\vec r})=\frac{1}{2}m\omega_0^2(y^2+z^2)-\hbar\omega_0$, where
the axial direction is equal to the $x$ axis (see
Fig.~\ref{trapped_gas}). 
Here we note that the Hamiltonian~(\ref{3DHamiltonian}) is invariant under 
the $Z_2$ reflection and the SO(2) rotation in the $y-z$ plane
[$\psi(x,y,z)\to\psi(x,-y,z)$ and 
$\psi(\vec r)\to \psi(x,y\cos\Phi -z\sin\Phi ,y\sin\Phi +z\cos\Phi )$, 
where $\Phi\in{\bol R}$]. Other symmetries of the
model~(\ref{3DHamiltonian}) are discussed in Appendix~\ref{sym_argument}.



\begin{figure}[tbp]
 \begin{center}
  \includegraphics[scale=0.4]{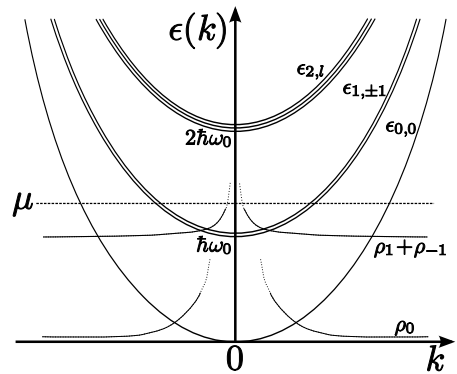}
  \caption{Single-particle energy band structure and density
  distributions (sketch) in each band. We assume that 
  the bosons live only in lower three energy bands. The
  density $\rho_\alpha$ is defined in Eq.~(\ref{S_tot}).}
  \label{bandstructure}
 \end{center}
\end{figure}

In order to make the one dimensionality of the
model~(\ref{3DHamiltonian}) more visible, 
it is useful to expand $\psi({\vec r})$ in terms of eigenstates of the 
2D harmonic oscillator $u_{n,l}(y,z)$ as follows:
\begin{equation}
 \psi({\vec r})=\sum_{n=0}^{\infty}\sum_{l}u_{n,l}(y,z)\phi_{n,l}(x),
 \label{expand}
\end{equation}
where $l$ is the angular momentum in the $y-z$ plane and can 
take $-n, -n+2,\dots, n-2,n$ for a given $n$. The angular momentum is a
good quantum number because of the SO(2) rotational symmetry of
Eq.~(\ref{3DHamiltonian}). The boson field $\phi_{n,l}$ describes 
the physics of the $x$ direction. Substituting Eq.~(\ref{expand}) 
into Eq.~(\ref{3DHamiltonian}), we obtain the single-particle energy
dispersion $\epsilon_{n,l}(k)=\frac{\hbar^2k^2}{2m}+\hbar\omega_0n$ 
which is shown in Fig.~\ref{bandstructure}.
It is noteworthy that all of the bands $\epsilon_{n,l}$ 
with the same index $n$ are degenerate, i.e., the $(n+1)$-fold
degeneracy exists. For example, as we already mentioned, 
the second lowest bands $\epsilon_{1,\pm 1}$ are doubly degenerate. 

To discuss effects of multi bands, we assume that the bosons occupy 
the lower three bands $\epsilon_{0,0}$, $\epsilon_{1,\pm 1}$ 
as in Fig.~\ref{bandstructure}. 
To satisfy this situation, we impose the following conditions: 
\begin{align}
 & \hbar\omega_0<\mu<2\hbar\omega_0,\label{chem} \\
 & a/a_{\perp}\ll 1 \hspace{1cm}({\rm i.e.,} \,\,\,
   \hbar\omega_0\gg U/a_{\perp}^3), \label{scatteringlength}\\
 &  2\hbar\omega_0-\mu\gg U/a_{\perp}^3,\label{scatteringlength2}
\end{align}
where the confinement radius $a_{\perp}=\sqrt{\hbar/m\omega_0}$
represents the width of the wave function $u_{n,l}(y,z)$ around the origin 
$(y,z)=(0,0)$. The first condition (\ref{chem}) ensures that 
the lower three bands have finite boson densities. The
other ones (\ref{scatteringlength}) and (\ref{scatteringlength2}) mean that 
the energy scale between neighboring bands is sufficiently larger than 
the strength of the interaction. 
These would allow us to neglect bosons in all of the higher bands
$\epsilon_{n>2,l}$ (i.e., bands $\epsilon_{n>2,l}$ are nearly empty of
bosons).

Under these assumptions, let us continue to study the
model~(\ref{3DHamiltonian}). 
Since the low-energy physics must be governed by the bosons on occupied
bands with $(n,l)=(0,0), (1,\pm 1)$, the low-energy effective theory can
be obtained by integrating out the degrees of freedom of the bosons on
all the vacant bands. In order to carry out this integration, it is
convenient to introduce the Euclidean action corresponding to 
the Hamiltonian~(\ref{3DHamiltonian}) through the path integral 
formalism as $S_{\rm tot}=S_0+S_1+S_{\rm int}$, 
\begin{subequations}
\label{S_tot}
\begin{align}
 S_0
 &=\int\!\! d\tau dx
    \sum_{n=0}^{1}\sum_{l}
     \phi_{n,l}^{*}
     \left[
      \partial_{\tau}
      -\frac{\hbar}{2m}\partial_x^2
      -\mu
      + n \hbar\omega_0
     \right]
     \phi_{n,l}
 \nonumber \\
 & 
   +\frac{U}{8\pi a_{\perp}}
    \Big[
     2\rho_{0,0}^2
     +\rho_{1,1}^2
     +\rho_{1,-1}^2
     +4\rho_{0,0}(\rho_{1,1}+\rho_{1,-1})\nonumber \\
  &  +4\rho_{1,1}\rho_{1,-1}
     +2(\phi_{0,0}^{*}\phi_{0,0}^{*}\phi_{1,1}\phi_{1,-1}+h.c.)
    \Big],
 \label{S0} \\
 S_1
 &=\int\!\! d\tau dx
    \sum_{n=2}^{\infty}\sum_{l}
     \phi_{n,l}^{\dagger}
     \left[
      \partial_{\tau}
      -\frac{\hbar}{2m}\partial_x^2
      -\mu
      + n \hbar\omega_0
     \right]
     \phi_{n,l},\label{S1}
  \\
 S_{\rm int}
 &=\frac{U}{2}
   \int\!\! d\tau dx
   \sideset{}{'}\sum_{\{n_i\}}
   \sideset{}{'}\sum_{\{l_i\}}
   P_{l_1,l_2,l_3,l_4}^{n_1,n_2,n_3,n_4}\nonumber\\
&  \hspace{3cm}\times\phi_{n_1,l_1}^{*}
   \phi_{n_2,l_2}^{*}
   \phi_{n_3,l_3}
   \phi_{n_4,l_4},\label{Sint}
\end{align}
\end{subequations}
where $\tau$ is the imaginary time, and 
$\rho_{n,l}=\phi_{n,l}^{*}\phi_{n,l}$ is a boson density 
of a band ($n,l$). The prime of the summations in $S_{\rm int}$ 
means the sum over all the possible sets 
of $(n_i,l_i)$ ($i=1,\cdots,4$) except for all of the interaction terms 
of $S_0$. Equations~(\ref{S0}) and (\ref{S1}) are, respectively, the
part including only the bosons of the occupied bands and the kinetic
part of massive bosons of all the empty bands. 
The remaining part $S_{\rm int}$ contains all the interaction terms 
including empty-band boson fields. The coupling constants 
$P_{l_1,l_2,l_3,l_4}^{n_1,n_2,n_3,n_4}$ in 
$S_{\rm int}$ are expressed as 
\begin{equation}
\label{coupling_const}
 P_{l_1,l_2,l_3,l_4}^{n_1,n_2,n_3,n_4}
 =\int\!\!dydz\,\, 
  u_{n_1,l_1}^*u_{n_2,l_2}^*u_{n_3,l_3}u_{n_4,l_4}.
\end{equation}


The integration of the empty-band bosons in $S_{\rm tot}$
can be performed by making use of 
the representations~(\ref{S0})-(\ref{Sint}) and the cumulant expansion. 
The resulting action is written as 
$S_{\rm 1D}=S_0+\langle S_{\rm int}\rangle_1
-\frac{1}{2}(\langle S_{\rm int}^2\rangle_1
-\langle S_{\rm int}\rangle_1^2)+\cdots$, 
where $\langle\cdots\rangle_1$ stands for the expectation value with
respect to $S_1$. For the action $S_{\rm 1D}$, effects of higher-energy bands
are taken into account through virtual scattering processes. 
In this calculation process, 
the Matsubara Green's function 
$G_n(\tau,x)=-\langle T_{\tau}\phi_{n,l}(x,\tau)\phi_{n,l}^*(0,0)\rangle$ of the
massive bosons with $\epsilon_{n>2,l}$ may be approximated as 
$\lambda_n^{-1}\Theta_s(\tau)\Theta_s(\lambda_n-|x|)\Theta_s(\Delta_n-|\tau|)$, 
where $\lambda_n=\sqrt{2\pi\hbar^2/m\Delta_n}$ is 
the thermal de Broglie wave length, 
$\Delta_n=n\hbar\omega_0-\mu$ is the gap of bosons, and 
$\Theta_s$ is the Heaviside's step function. 

The resulting effective Hamiltonian corresponding to the action $S_{\rm 1D}$ 
is given as 
\begin{subequations}
\label{1D_Hamiltonian_all}
\begin{align}
 H_{1D}
 &= \int\!\! dx\,\,\,{\cal H}_0+{\cal H}_{\rm int},
     \label{1D_Hamiltonian} \\
 {\cal H}_0
 &= \frac{\hbar^2}{2m}\sum_{\alpha}
    \partial_x\phi^{\dagger}_{\alpha}\partial_x\phi_{\alpha}
    -\mu_0\rho_0-\mu_1(\rho_1+\rho_{-1}),
    \label{1D_Hamiltonian_1} \\
 {\cal H}_{\rm int}
 &= \frac{U}{8\pi a_{\perp}^2}
    \Big[
     2\Gamma_0\rho_0^2+\Gamma_1(\rho_1^2+\rho_{-1}^2)
     +4\Gamma_{01}\rho_0(\rho_1+\rho_{-1})
    \nonumber \\
 &  
     +4\Gamma_{\pm1}\rho_{1}\rho_{-1}
     +2\Gamma_{0\pm1}
     (
      \phi_{0}^{\dagger}\phi_{0}^{\dagger}\phi_{1}\phi_{-1}+h.c.
     )
    \Big], \label{1D_Hamiltonian_int} 
\end{align}
\end{subequations}
where $\alpha=0, \pm1$ correspond to band indices 
$(n,l)=(0,0), (1,\pm1)$, respectively. 
The Hamiltonian ${\cal H}_0$ is the free boson part and $\mu_0=\mu>0$ and 
$\mu_1=\mu-\hbar\omega_0>0$ are the chemical potentials. 
The first and second terms in ${\cal H}_{\rm int}$ are the two-body
interactions in the same band. The third and fourth are those 
between the different bands. The final term is the tunneling 
between the lowest band and the doubly degenerate bands. 
The coupling constants $\Gamma_\alpha$ are modified from 
unity (the bare value in $S_0$) due to vacant-band effects. 

Calculating lower-order cumulants carefully, one can find that, for
example, massive bosons on $\epsilon_{2,l}$ varies coupling constants as 
$\Gamma_{\alpha}\approx 1+\sum_{n=1}^\infty c_{\alpha,n}(a/a_\perp)^n$
($c_{\alpha,n}$ are nonuniversal constants). This result strongly
supports that the cumulant expansion is reliable and works out 
under the conditions~(\ref{chem})-(\ref{scatteringlength2}) and 
$\Gamma_\alpha=1$ is semi-quantitatively correct. 
The cumulant expansion generates other three- or more-body 
interactions, but they are also negligible compared with the 
two-body ones in ${\cal H}_{\rm int}$.


\section{Analysis I}\label{analysis1}
In the preceding section, we have obtained the 1D effective
Hamiltonian~(\ref{1D_Hamiltonian_all}). Here, analyzing it, let us
investigate the low-energy properties of the Bose gas under the
conditions~(\ref{chem})-(\ref{scatteringlength2}). 
First, to evaluate the mean density profile of each band, we 
introduce the following GL potential 
from ${\cal H}_{\rm int}$,
\begin{eqnarray}
 {\cal F}_{\rm GL}
 = -\mu_0\rho_{0}-\mu_1\rho_{s}
    +\frac{U}{8\pi a_{\perp}^2}
     \big[
      2\Gamma_0\rho_0^2+\Gamma_s\rho_s^2
      +\Gamma_{a}\rho_{a}^2     
     \nonumber \\
   +4\Gamma_{01}\rho_{0}\rho_s
     +2\Gamma_{0\pm1}\rho_0\sqrt{\rho_s^2-\rho_a^2}\cos(2\theta_0-2\theta_s)
     \big], \label{GLpotential}
\end{eqnarray}
where $\Gamma_{s,a}=(\Gamma_{1}\pm2\Gamma_{\pm1})/2$ and 
we have defined $\phi_{\alpha}=\rho_{\alpha}^{1/2}e^{i\theta_{\alpha}}$, 
$\rho_{s,a}=\rho_{1}\pm\rho_{-1}$ and 
$\theta_{s,a}=(\theta_{1}\pm\theta_{-1})/2$. 
The quantity $\rho_s$ is the total density in the second lowest bands, 
while $\rho_a$ stands for the angular momentum density in the same bands 
[the total angular momentum is 
${\cal L}=\sum_{n,l}\int\!\! dx\ l\rho_{n,l}(x)$]. 
Therefore, a solution with $\rho_a\neq 0$, if possible, 
means the emergence of the spontaneous rotation and 
the breakdown of the $Z_2$ reflection symmetry in the trapped plane. 
Each term in ${\cal F}_{\rm GL}$ has the following physical meanings: 
(\rnum{1}) Positive $\mu_{0,1}$ induce finite densities 
$\rho_{0,\pm1}\neq 0$ and the relation $\mu_0\gg\mu_1$ 
must yield $\rho_0\gg \rho_{\pm1}$, 
(\rnum{2}) $\Gamma_{0,s}(>0)$ tend to decrease $\rho_{0,s}$,
while $\Gamma_a(<0)$ promotes the growth of $\rho_a$, 
(\rnum{3}) $\Gamma_{01}(>0)$ favors the decrease of $\rho_{0,s}$,
and (\rnum{4}) the tunneling $\Gamma_{0\pm1}(>0)$ inversely enhances
$\rho_{0,s}$ and decreases $\rho_a$.
Recalling the condition~(\ref{scatteringlength}), we
can guess that bosons on the second lowest bands interact only 
with the lowest-band bosons with a kinetic energy 
$\epsilon_{0,0}(k)\sim\hbar\omega_0$, and hardly 
influence the bosons with lower energies. 
It is hence expected that $\Gamma_{01}$ and $\Gamma_{0\pm1}$ are 
overestimated in ${\cal F}_{\rm GL}$. 
Here, we simply regard both the density 
amplitude $\rho_{\alpha}$ and the phase $\theta_{\alpha}$ as $c$
numbers, 
although for real quantum systems, we cannot fix them 
simultaneously. This approximation would also overestimate 
the effects of $\Gamma_{0\pm1}$. 

The GL equations
$\partial{\cal F}_{\rm GL}/\partial\rho_{\alpha}
=\partial{\cal F}_{\rm GL}/\partial\theta_{\alpha}=0$
are solved as
\begin{subequations}
\label{MF_analysis1} 
\begin{align}
 & {\bar \rho}_0
 = \frac{4\pi a_{\perp}^2}{U}
     \frac{\Gamma_s\mu_0-(2\Gamma_{01}-2\Gamma_{0\pm1})\mu_{1}}
          {2\Gamma_0\Gamma_s-(2\Gamma_{01}-\Gamma_{0\pm1})^2},
   \label{MFS1}  \\
 & {\bar \rho}_s
 = \frac{4\pi a_{\perp}^2}{U}
     \frac{-(2\Gamma_{01}-\Gamma_{0\pm1})\mu_{0}+2\Gamma_{0}\mu_{1}}
          {2\Gamma_0\Gamma_s-(2\Gamma_{01}-\Gamma_{0\pm1})^2},
   \label{MFS2}  \\
 & {\bar \rho}_a = 0, \label{MFS3} \\
 & {\bar \theta}_0-{\bar\theta}_s =\pi/2. \label{MFS4}
\end{align}
\end{subequations}
Furthermore, computing 
$\partial^2{\cal F}_{\rm GL}/\partial\rho_\alpha\partial\rho_\beta$, we 
find that the above solution is stable and minimize ${\cal F}_{\rm GL}$. 
Because of $\bar \rho_a=0$, the GL argument shows no
symmetry breaking. If we straightforwardly adopt the approximation 
$\Gamma_\alpha=1$, $\bar\rho_s$ becomes negative 
under the condition~(\ref{chem}). This is unphysical and attributed 
to large $\Gamma_{01}$ and $\Gamma_{0\pm1}$, as expected. 
To recover a physical solution $\bar\rho_0>\bar\rho_s\neq 0$, 
we should replace $\Gamma_{01}$ and $\Gamma_{0\pm1}$ with small values, 
$\tilde\Gamma_{01}$ and $\tilde\Gamma_{0\pm1}$, respectively: For example, 
when we set $\tilde\Gamma_{01}=\tilde\Gamma_{0\pm1}=a/a_{\perp}\ll1$, the GL
equations offer a reasonable solution 
$\bar\rho_s\approx\frac{8\pi a_{\perp}^2}{3U}
(\mu_1-\frac{a}{2a_{\perp}}\mu_0)$.

Let us next take into account the quantum fluctuation around the
physical GL solution. To this end, 
the bosonization~\cite{Haldane,Cazalilla,Giamarchi} is powerful and
unbiased. It makes the density and field operators transform as
\begin{subequations}
\label{bosonization0}
\begin{align}
 \rho_{\alpha}
 &\approx
  \left(
   \bar{\rho}_{\alpha}+\frac{1}{\pi}\partial_x\varphi_{\alpha}
  \right)
  \sum_{n=-\infty}^{\infty}
   e^{i2n(\varphi_{\alpha}-\pi\bar{\rho}_{\alpha}x)},
 \label{bosonization1}\\
 \phi_{\alpha}
 &\approx
  \sqrt{\bar{\rho}_{\alpha}+\frac{1}{\pi}\partial_x\varphi_{\alpha}}
  \sum_{n=-\infty}^{\infty}
   e^{i2n(\varphi_{\alpha}-\pi\bar{\rho}_{\alpha}x)}
   e^{-i({\bar\theta}_{\alpha}+\theta_{\alpha})},
 \label{bosonization2}
\end{align}
\end{subequations}
where the phase fields $\theta_{\alpha}$ and $\varphi_{\alpha'}$
($\alpha,\alpha'=0,\pm 1$) 
represent the quantum fluctuation and obey 
$[\theta_{\alpha}(x),\partial_{x'}\varphi_{\alpha'}(x')]
=[\varphi_{\alpha}(x),\partial_{x'}\theta_{\alpha'}(x')]
=i\pi\delta_{\alpha,\alpha'}\delta(x-x')$.
Here we define $\varphi_{s,a}=\varphi_{1}\pm\varphi_{-1}$.
Substituting the formula (\ref{bosonization0}) 
into Eq.~(\ref{1D_Hamiltonian_all}), 
we obtain the phase-field Hamiltonian, which is invariant under the
symmetry operations~(\ref{symm_operation}) and (\ref{symm_operation2}). 
Remarkably, due to $\Gamma_a<0$, the coefficient of 
$(\partial_x\varphi_a)^2$ becomes negative in the Hamiltonian. 
This means that the ($\varphi_a$, $\theta_a$) sector has an 
instability against the fluctuation of $\rho_a$. 
It might be restored by sufficiently large higher-order 
differential terms such as $(\partial_x\varphi_a)^4$~\cite{Yang}, 
but the emergence of those terms
would not be expected under the condition~(\ref{scatteringlength}). 
We thus conclude that the quantum fluctuation violates the GL
solution and induces a ground state with a finite angular momentum
$\bar\rho_a\neq 0$. It is known that similar scenarios of symmetry
breakings can occurs in a few systems.~\cite{CH,KV,Yang,SS}

\section{Analysis II}\label{analysis2}
In order to examine and to enhance the validity of 
the above prediction of $\bar\rho_a\neq 0$, let us reconsider the
effective theory~(\ref{1D_Hamiltonian_all}) using another approximation. 
As we already mentioned, $\Gamma_{01}$ and $\Gamma_{0\pm1}$ terms are
expected to be overestimated in Eq.~(\ref{GLpotential}). Therefore, at first
we daringly drop the tunneling term with $\Gamma_{0\pm1}$ and 
replace $\Gamma_{01}$ with a small effective value $\tilde\Gamma_{01}$. 
For this simplified case, the GL potential is written as 
\begin{eqnarray}
 \tilde{\cal F}_{GL}
 & =& -\mu_0\rho_0-\mu_1\rho_s
     +\frac{U}{8\pi a_{\perp}^2}
       \big[
        2\Gamma_0\rho_0^2+\Gamma_s\rho_s^2
     \nonumber \\
    &&   +\Gamma_a\rho_a^2
        +4\tilde\Gamma_{01}\rho_0\rho_s
       \big]. \label{GLpotential2}
\end{eqnarray}
The GL equations $\partial\tilde{\cal F}_{GL}/\partial\rho_{\alpha}=0$
lead to ${\bar \rho}_0=\frac{2\pi a_{\perp}^2}{U}
\frac{\Gamma_s\mu_0-2\tilde\Gamma_{01}\mu_1}
{\Gamma_0\Gamma_s-2\tilde\Gamma_{01}}$,
${\bar \rho}_s=\frac{4\pi a_{\perp}^2}{U}
\frac{\Gamma_0\mu_1-\tilde\Gamma_{01}\mu_0}
{\Gamma_0\Gamma_s-2\tilde\Gamma_{01}}$ and ${\bar \rho}_a=0$. 
Namely, the GL solution again suggests no angular momentum. 
However, considering the stability of the solution, 
we see that the density profile $(\bar\rho_0,\bar\rho_s,\bar\rho_a)$
corresponds to the saddle point for $\tilde{\cal F}_{GL}$ 
and it is minimized at $\rho_a\to\pm\infty$ which is unphysical. 
The reason for this instability is that the interband interaction
$\Gamma_{\pm1}$ is stronger than intraband one $\Gamma_{1}$.  
To recover a physically proper solution, we can
add the following phenomenological term 
$\frac{U}{8\pi a_\perp^2}\frac{\Gamma_a^{(4)}}{4}\rho_a^4$  
$(\Gamma_a^{(4)}>0)$
to $\tilde{\cal F}_{GL}$.~\cite{comment1}
It is interpreted that this term originates from the higher-order
corrections of the cumulant expansion or many-body
interactions neglected in the original Hamiltonian~(\ref{3DHamiltonian}). 
This modification offers a stable, physical solution
\begin{equation}
{\bar \rho}_a=\pm\sqrt{-\frac{2\Gamma_a}{\Gamma_a^{(4)}}}. \label{MFS5}
\end{equation}
From the consideration above, we see that the GL approach with
neglecting the tunneling term naturally 
leads to a finite angular momentum.

As in the previous analysis in Sec.~\ref{analysis1}, 
let us bosonize the Hamiltonian~(\ref{1D_Hamiltonian_all}) 
based on the GL solution (\ref{MFS5}). 
At this stage, we recover the tunneling term. 
The resulting phase-field Hamiltonian is represented as 
\begin{eqnarray}
 {\cal H}_{1D}
 &\approx&
 \int\!\!dx
   \sum_{\alpha=0,s,a}
   \frac{v_{\alpha}}{2\pi}
    \left\{
     K_{\alpha}(\partial_x\theta_{\alpha})^2
     +K_{\alpha}^{-1}(\partial_x\varphi_{\alpha})^2
   \right\}
 \nonumber \\
 && \hspace{1.5cm}  +\,\,{\tilde g}_{sa}\partial_x\theta_s\partial_x\theta_a
    +{\tilde g}_{0s}\partial_x\varphi_0\partial_x\varphi_s
 \nonumber \\
 &&  \hspace{1.5cm} +\,\,{\tilde g}\cos(2\theta_0-2\theta_s)+\cdots.
 \label{PhaseHamiltonian2}
\end{eqnarray}
Coefficients of $(\partial_x\theta_0)^2$, $(\partial_x\theta_s)^2$ and 
$(\partial_x\theta_a)^2$ are proportional to $\bar\rho_0$, $\bar\rho_s$
and $\bar\rho_s$, respectively, while 
${\tilde g}_{sa}\propto{\bar \rho}_a$.

Therefore, when ${\bar \rho}_a$ is smaller enough than 
$\bar\rho_{0,s}$, 
any instability of restoring ${\bar \rho}_a=0$
does not originate from the differential terms. 
In addition, the symmetry operations~(\ref{symm_operation2}) indicate
that any vertex term with $\varphi_a$ or $\theta_a$ is forbidden in the
Hamiltonian~(\ref{PhaseHamiltonian2}).~\cite{SS}
Thus, we can say that $\bar\rho_a\ne 0$ 
is not destroyed by the quantum fluctuation and 
it agrees with our previous approach in Sec.~\ref{analysis1}.
The symmetries also tell us that for all of the vertex operators, 
only $\cos[2n(\theta_{0}-\theta_{s})]$ ($n\in {\bol Z}$) are
allowed to appear in the phase-field Hamiltonian. 
The final term of Eq.~(\ref{PhaseHamiltonian2}) is the most relevant in
the allowed vertex interactions.
Generally, for 1D boson systems with repulsive short-range interactions, 
the TLL parameter $K_{0,s,a}$ are known to be larger than
unity.~\cite{Cazalilla} The vertex term $\cos(2\theta_0-2\theta_s)$ 
thus must be strongly relevant~\cite{Giamarchi} and 
$\theta_0-\theta_s$ is locked at $\pi/2$, which is consistent 
with our previous result~(\ref{MFS4}). 
Introducing new fields $(\varphi_\pm,\theta_\pm)
=(\varphi_0\pm\varphi_s,\theta_0\pm\theta_s)/\sqrt{2}$,
we can interpret that the cosine term opens a gap in the
$(\varphi_-,\theta_-)$ sector. Following the method in
Ref.~\cite{BF}, we can trace out the degrees of freedom 
of the gapped sector. After that, we finally obtain a 
two-component TLL Hamiltonian with the $(\varphi_{+},\theta_{+})$
and $(\varphi_{a},\theta_{a})$ sectors as the low-energy effective
theory.

On the other hand, when $\bar\rho_a$ is too large in Eq.~(\ref{MFS5}), 
we should adopt $(\bar\rho_1,\bar\rho_{-1})=(\bar\rho_s,0)$ or 
$(0,\bar\rho_s)$ as the proper GL solution instead of Eq.~(\ref{MFS5}). 
For this case, neglecting all of the parts with the field $\phi_{-1}$ in
the Hamiltonian~(\ref{1D_Hamiltonian_all}), 
we again derive a two-component TLL Hamiltonian.

Consequently, we conclude from the second analysis
above that the ground state possesses a finite angular momentum 
$\bar\rho_a\neq 0$ and the low-energy excitations are described 
by a two-component TLL.~\cite{comment2} 

\begin{figure}[tbp]
 \begin{center}
  \includegraphics[scale=0.45]{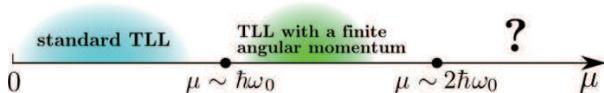}
  \caption{Predicted ground-state phase diagram of the
  model~(\ref{3DHamiltonian}) under the
  condition~(\ref{scatteringlength}). 
  Quantum phase transitions are expected to
  occur around the black points. They, however, might be interrupted as
  a crossover phenomenon rather than a transition, since boson densities
  of higher bands would be nonzero even when the chemical potential $\mu$
  is smaller than $\hbar\omega_0$. When $\mu$ is larger than
  $n\hbar\omega_0$ ($n>2$), the bosons on higher bands
  $\epsilon_{n>2,l}$ would also possess an angular momentum. It is
  therefore expected that, for example, the bosons on $\epsilon_{2,l}$
  rotate clockwise, while those on $\epsilon_{3,l}$ do
  anticlockwise. Namely, the almost complete or partial cancellation of
  the angular momentum can occur.
  } 
\label{PhaseDiagram}
 \end{center}
\end{figure}

\section{Summary and discussions}\label{summary}

In this paper, we have studied the 1D harmonically trapped Bose
gas~(\ref{3DHamiltonian}) with bosons on multi-transverse
modes. Starting from the 3D boson systems with 2D confinement potential,
the 1D effective Hamiltonian is derived by integrating out the degrees
of freedom of all the vacant bands. 
Applying two theoretical ways based on the GL approach and the
bosonization to the 1D effective Hamiltonian~(\ref{1D_Hamiltonian_all}), 
we have shown that when bosons are filled in the lower
three bands under the conditions~(\ref{chem})-(\ref{scatteringlength2}),
a angular momentum becomes finite and the $Z_2$ reflection symmetry in
the trapped plane is spontaneously broken. This phenomenon essentially
originates from the larger
repulsive interaction between the degenerate bands $(n,l)=(1,\pm 1)$ than
the intraband repulsion. 
The second analysis in Sec.~\ref{analysis2} predicts 
that the low-energy excitations on the rotating ground state 
are described by a two-component TLL.~\cite{comment2} 
It is known that when the bosons occupy only the lowest bands, 
the low-energy physics is governed by a one-component TLL. 
We thus can draw the phase diagram in Fig.~\ref{PhaseDiagram}.

The conjugate field for $\rho_a$ is not realistic in the
model~(\ref{3DHamiltonian}). Any trigger of the rotating ground state
therefore seems to be absent. However, the real trap potential must
deviate from the harmonic type. Such a small deviation would help the
system circulate clockwise or anticlockwise. Thermal fluctuations 
and large deformations of the trapped potential would generate kink
structures between clockwise and anticlockwise rotating regimes.
When Bose atoms carry an electric charge, the predicted rotation means
spontaneous loop current and magnetic flux. If we suddenly change 
the form of the trap potential and then observe the real-space boson 
density profile, we can detect, in principle, whether the bosons 
rotate clockwise or anticlockwise.

\appendix
\section{Symmetry argument}
\label{sym_argument}
Here, we briefly summarized the symmetries of our Bose gas
system~(\ref{3DHamiltonian}). They are often used in our analyses of 
Secs.~\ref{analysis1} and \ref{analysis2}. 

The Hamiltonian (\ref{3DHamiltonian}) is
invariant under the following symmetry operations: 
The translation and the parity for the $x$-axis
[$\psi(x,y,z)\to\psi(x+\delta,y,z)$ ($\delta\in {\bol R}$) and 
$\psi(x,y,z)\to\psi(-x,y,z)$],
the $Z_2$ reflection and the SO(2) rotation in the $y$-$z$ plane
[$\psi(x,y,z)\to\psi(x,-y,z)$ and 
$\psi(\vec r)\to \psi(x,y\cos\Phi -z \sin\Phi ,y\sin\Phi +z\cos\Phi )$ 
($\Phi\in{\bol R}$)], 
and the global U(1) gauge transformation
[$\psi(x,y,z)\to\psi(x,y,z)e^{i\Theta}$ ($\Theta\in{\bol R}$)].

For the bosonized phase-field Hamiltonians (e.g.,
Eq.~(\ref{PhaseHamiltonian2})), it is important to note that all of the 
above symmetry operations can be translated to the bosonization
language.~\cite{SS} The translation and the parity operation for the
$x$-direction, the $Z_2$ reflection and the SO(2) rotation in the
$y$-$z$ plane, and the global U(1) gauge transformation are,
respectively, expressed as 
\begin{subequations}
\label{symm_operation}
\begin{align}
 & [\varphi_{\alpha}(x),\theta_{\alpha}(x)]
   \rightarrow 
   [\varphi_{\alpha}(x+\delta)
   +\pi\bar{\rho}_{\alpha}\delta,\theta_{\alpha}(x+\delta)], 
 \label{operation1}\\
 & [\varphi_{\alpha}(x),\theta_{\alpha}(x)]
   \rightarrow 
   [-\varphi_{\alpha}(-x),\theta_{\alpha}(-x)],
 \label{operation2}\\
 & [\varphi_{\alpha}(x),\theta_{\alpha}(x),\bar\rho_{\alpha}]
   \rightarrow 
   [\varphi_{-\alpha}(x),\theta_{-\alpha}(x),\bar\rho_{-\alpha}],
 \label{operation3}\\
 & [\varphi_{\alpha}(x),\theta_{\alpha}(x)]
   \rightarrow 
   [\varphi_{\alpha}(x),\theta_{\alpha}(x)-\alpha\Phi],
 \label{operation4}\\
 & [\varphi_{\alpha}(x),\theta_{\alpha}(x)]
   \rightarrow 
   [\varphi_{\alpha}(x),\theta_{\alpha}(x)+\Theta],
 \label{operation5}
\end{align}
\end{subequations}
where $\alpha=0,\pm1$. Using another set of the phase fields 
$(\varphi_{s,a},\theta_{s,a})=(\varphi_{1}\pm\varphi_{-1},\frac{\theta_{1}\pm\theta_{-1}}{2})$,
we can further transform Eq.~(\ref{symm_operation}) as 
\begin{subequations}
\label{symm_operation2}
\begin{align}
 & [\varphi_a(x),\theta_a(x)]
   \rightarrow 
   [\varphi_a(x+\delta)+\pi\bar{\rho}_a\delta,\theta_a(x+\delta)],
 \label{operation2-1}\\
 & [\varphi_a(x),\theta_a(x)]
   \rightarrow 
   [-\varphi_a(-x),\theta_a(-x)],
 \label{operation2-2} \\
 & [\varphi_a(x),\theta_a(x),\bar\rho_a]
   \rightarrow 
   [-\varphi_a(x),-\theta_a(x),-\bar\rho_a],
 \label{operation2-3}\\
 & [\varphi_a(x),\theta_a(x)]
   \rightarrow 
   [\varphi_a(x),\theta_a(x)-\Phi],
 \label{operation2-4}\\
 & [\varphi_a(x),\theta_a(x)]
   \rightarrow 
   [\varphi_a(x),\theta_a(x)],
 \label{operation2-5}
\end{align}
\end{subequations}
for $(\varphi_a,\theta_a)$, and
\begin{subequations}
\label{operation3}
\begin{align}
 & [\varphi_s(x),\theta_s(x)]
   \rightarrow 
   [\theta_s(x+\delta),\varphi_s(x+\delta)+\pi\bar{\rho}_s\delta],
 \label{operation3-1}\\
 & [\varphi_s(x),\theta_s(x)]
   \rightarrow 
   [-\varphi_s(-x),\theta_s(-x)],
 \label{operation3-2}\\ 
 & [\varphi_s(x),\theta_s(x)]
   \rightarrow 
   [\varphi_s(x),\theta_s(x)],
 \label{operation3-3}\\ 
 & [\varphi_s(x),\theta_s(x)]
   \rightarrow 
   [\varphi_s(x),\theta_s(x)],
 \label{operation3-4}\\ 
 & [\varphi_s(x),\theta_s(x)]
   \rightarrow 
   [\varphi_s(x),\theta_s(x)+\Theta], 
 \label{operation3-5}
\end{align} 
\end{subequations}
for $(\varphi_s,\theta_s)$. Note that $\bar\rho_a$ changes the sign
under the $Z_2$ reflection in the $y$-$z$ plane shown in
Eq.~(\ref{operation2-3}). 

For instance, applying these symmetries, we can strongly restrict 
possible operators in the bosonized Hamiltonian~(\ref{PhaseHamiltonian2}).

\begin{acknowledgments}
The authors are grateful to M. Cazalilla, K. Kamide, M. Ueda,
A. Furusaki, and M. Oshikawa for useful discussions. 
A. T. was supported by JSPS and M. S. was supported 
by Grant-in-Aid for Scientific Research from MEXT (Grant 
No. 17071011). 
\end{acknowledgments}


\begin{thebibliography}{}
\bibitem{comment1}
Actually, $\Gamma_a^{(4)}$ is not always positive. If
$\Gamma_a^{(4)}$ is negative, the further higher order
correction is needed.
\bibitem{comment2} 
If $\tilde g_{0s}\propto 4\tilde\Gamma_{01}$ is comparable with 
coefficients of $(\partial_x\varphi_{0,s})^2$ or $\bar \rho_a$ 
approaches $\bar\rho_s$ in Eq.~(\ref{PhaseHamiltonian2}), 
the low-energy excitation structure 
might qualitatively change.~\cite{Yang}
\end{thebibliography}

\end{document}